\documentclass[12pt]{aastex631}
\usepackage{natbib}
\usepackage{amsmath}
\usepackage{verbatim}
\usepackage{amssymb}
\usepackage{hyperref}
\usepackage{float}
\usepackage{graphicx}
\usepackage{mathtools}
\usepackage{empheq}
\usepackage{appendix}
\usepackage{microtype}
\usepackage{commath}
\usepackage{bm}
\usepackage{bold-extra}

\DeclareGraphicsExtensions{.pdf,.png,.jpg}

\newcommand{\kmsmpc}{\hbox{$ \, \rm km\, s^{-1} \, Mpc^{-1}$}}

\newcommand{\bq}{\begin{equation}} 
\newcommand{\eq}{\end{equation}} 
 
\newcommand{\kms}{km s$^{-1}$ Mpc$^{-1}$}

\newcommand{\Gaia}{{\it Gaia\ }}

\newcommand{\beq}{\begin{equation}}
\newcommand{\eeq}{\end{equation}}
\newcommand{\beqa}{\begin{eqnarray}}
\newcommand{\eeqa}{\end{eqnarray}}

\newcommand{\PL}{$P$--$L$\ }

\shorttitle{Cluster Cepheids with \Gaia and {\it HST}}
\shortauthors{Riess et al.}

\begin{document} 

\title{Cluster Cepheids with High Precision \Gaia Parallaxes, Low Zeropoint Uncertainties,\\and {\it Hubble Space Telescope} Photometry} 

\author[0000-0002-6124-1196]{Adam G.~Riess}
\affiliation{Space Telescope Science Institute, 3700 San Martin Drive, Baltimore, MD 21218, USA}
\affiliation{Department of Physics and Astronomy, Johns Hopkins University, Baltimore, MD 21218, USA}

\author[0000-0003-3889-7709]{Louise Breuval}
\affiliation{Department of Physics and Astronomy, Johns Hopkins University, Baltimore, MD 21218, USA}

\author[0000-0001-9420-6525]{Wenlong Yuan}
\affiliation{Department of Physics and Astronomy, Johns Hopkins University, Baltimore, MD 21218, USA}

\author{Stefano Casertano}
\affiliation{Space Telescope Science Institute, 3700 San Martin Drive, Baltimore, MD 21218, USA}

\author[0000-0002-1775-4859]{Lucas M.~Macri}
\affiliation{George P.\ and Cynthia W.\ Mitchell Institute for Fundamental Physics and Astronomy,\\ Department of Physics \& Astronomy, Texas A\&M University, College Station, TX 77843, USA}

\author{J.~Bradley Bowers}
\affiliation{Department of Physics and Astronomy, Johns Hopkins University, Baltimore, MD 21218, USA}

\author[0000-0002-4934-5849]{Dan Scolnic}
\affiliation{Department of Physics, Duke University, Durham, NC 27708, USA}

\author[0000-0002-4934-5849]{Tristan Cantat-Gaudin}
\affiliation{Max-Planck-Institut f{\"u}r Astronomie, K{\"o}nigstuhl 17, D-69117
Heidelberg, Germany}

\author[0000-0001-8089-4419]{Richard I. Anderson}
\affiliation{Institute of Physics, Laboratory of Astrophysics, \'Ecole Polytechnique F\'ed\'erale de Lausanne (EPFL), Observatoire de Sauverny, 1290 Versoix, Switzerland}

\author[0000-0003-2443-173X]{Mauricio Cruz Reyes}
\affiliation{Institute of Physics, Laboratory of Astrophysics, \'Ecole Polytechnique F\'ed\'erale de Lausanne (EPFL), Observatoire de Sauverny, 1290 Versoix, Switzerland}

\begin{abstract}

We present {\it HST} photometry of 17 Cepheids in open clusters and their cluster mean parallaxes from \Gaia EDR3. These parallaxes are more precise than those from individual Cepheids ($G<8$ mag) previously used to measure the Hubble constant because they are derived from an average of $>$ 300 stars per cluster. Cluster parallaxes also have smaller systematic uncertainty because their stars lie in the range ($G>13$ mag) where the \Gaia parallax calibration is the most comprehensive. Cepheid photometry employed in the period--luminosity relation was measured using the same {\it HST} instrument (WFC3) and filters ({\it F555W}, {\it F814W}, {\it F160W}) as extragalactic Cepheids in Type Ia supernova (SN Ia) hosts. We find no evidence of residual parallax offset in this magnitude range, $zp=-3 \pm$4 $\mu$as, consistent with the results from \cite{Lindegren:2021b} and most studies. The Cepheid luminosity (at $P\,=\,10$~d and solar metallicity) in the {\it HST} near-infrared, Wesenheit magnitude system derived from the cluster sample is $M_{H,1}^W=-5.902 \pm 0.025$ mag and $-5.890 \pm 0.018$ mag with or without simultaneous determination of a parallax offset, respectively. These results are similar to measurements from field Cepheids, confirming the accuracy of the \Gaia parallaxes over a broad range of magnitudes. The SH0ES distance ladder calibrated only from this sample gives $H_0=72.9 \pm 1.3$ and $H_0=73.3 \pm 1.1$ \kms\ with or without offset marginalization; combined with all other anchors we find $H_0=73.01 \pm 0.99$ and $73.15 \pm 0.97$ \kms, respectively, a 5\% or 7\% reduction in the uncertainty in $H_0$ and a $\sim5.3\sigma$ Hubble Tension relative to Planck+$\Lambda$CDM. It appears increasingly difficult to reconcile two of the best measured cosmic scales, parallaxes from \Gaia and the angular size of the acoustic scale of the CMB, using the simplest form of $\Lambda$CDM to connect the two. 

\end{abstract}

\clearpage

\section{Introduction}

Trigonometric parallaxes of Milky Way Cepheids measured by the {\it ESA Gaia} mission, in concert with their fluxes measured with the Hubble Space Telescope {\it (HST}), offer the only route at present to reach a 1\% geometric calibration of a distance ladder used to measure the Hubble constant. The \Gaia EDR3 data release provided parallaxes with $\sim$25 $\mu$as precision for 75 Cepheids photometrically with {\it HST} to reach a 1.1\% geometric calibration including marginalization over the \Gaia parallax offset term \citep[][hereafter R21]{Riess:2021}.  These were used in the recent SH0ES measurement of $H_0$ 
\citep[][hereafter R22]{Riess:2022} and in other studies \citep{Broutetal:2022,Jones:2022}. However, it is possible to obtain parallaxes with still greater individual precision and lower systematic uncertainty for Cepheids that reside in Milky Way open clusters \citep{Turner:2010}.

The precision of the mean parallax of a cluster of stars can greatly exceed that of its individual members. \citet[][hereafter B20]{Breuval:2020} confirmed this for 13 Milky Way cluster Cepheids with parallaxes from \Gaia DR2. Although cluster parallax uncertainties do not scale as the square root of the number of stars in a Cepheid-hosted cluster due to the angular covariance of \Gaia parallax measurements \citep[][hereafter L21]{Lindegren:2021b}, a mean cluster parallax uncertainty including the angular uncertainty is approximately three times smaller (on average) than that of the Cepheid it hosts and thus each cluster Cepheid has the weight of approximately nine field Cepheids.

However, Cepheids are rarely found in star clusters. Although born in young clusters, their ages ($\sim$ 100 Myr) greatly exceed the dissipation timescale of open clusters \citep[$\sim$ 10 Myr,][]{Lada:2003,Dinnbier:2022}, so that Cepheids hosted by clusters comprise only a few percent of the Cepheid population \citep{Anderson:2013}. Nevertheless, their unique leverage makes even a modest sample significant and valuable.

\begin{figure}[b] 
\begin{center}
\includegraphics[width=0.6\textwidth]{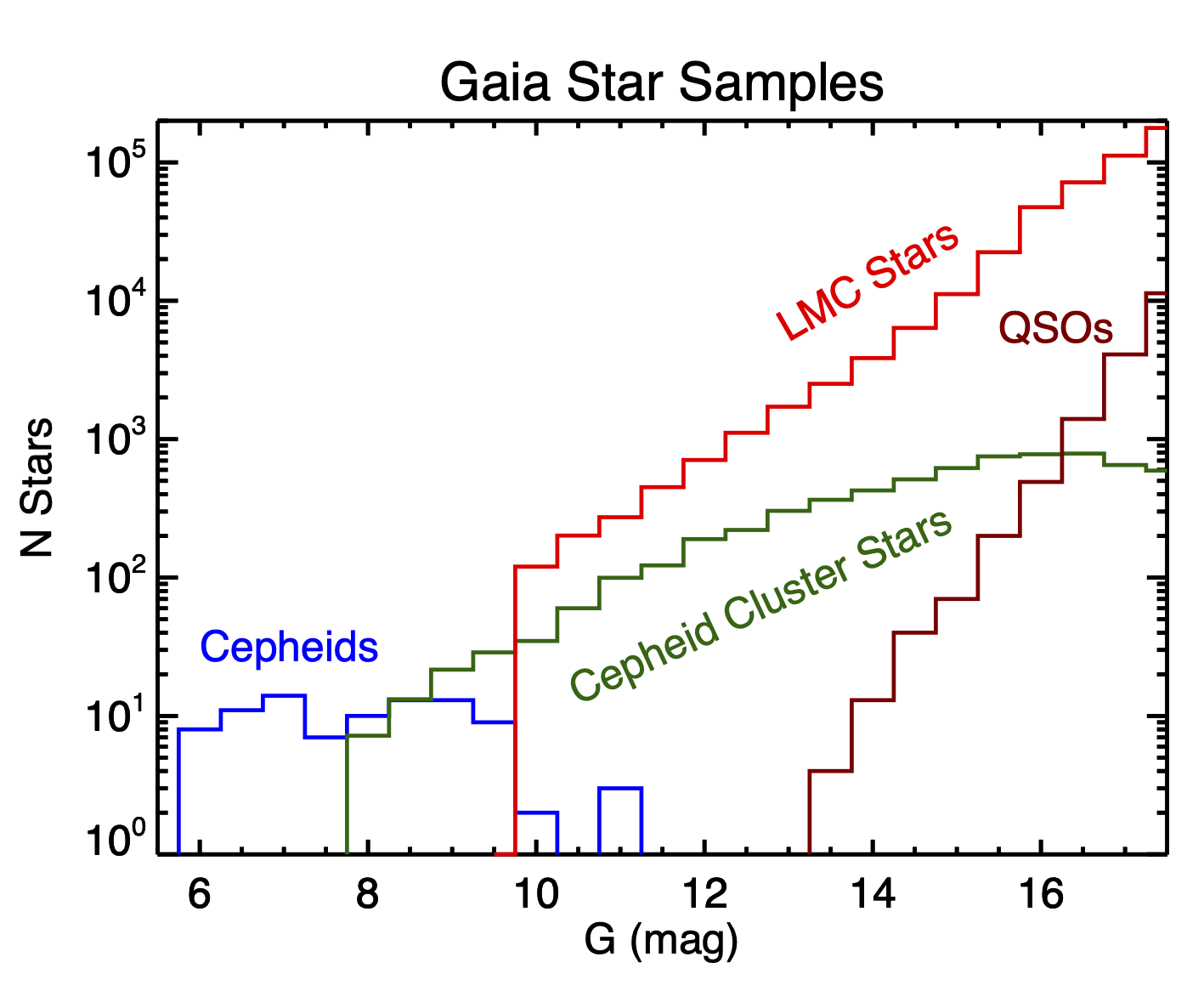}
\end{center}
\caption{\label{fg:hist} Sources used by L21 to calibrate the \Gaia EDR3 parallax offset (LMC stars and QSOs) and samples used to measure the parallaxes of Cepheids. L21 also used $\sim$ 7000 physical pairs of stars (not in this plot) with the brighter member having $6 < G < 10$ to calibrate the bright range. The parallaxes of field Cepheids (blue line) used in R21 and the SH0ES determination of $H_0$ require the simultaneous determination of the parallax offset due to the more limited \Gaia calibration in this range. The parallaxes of Cepheids in clusters were determined with a far greater number of stars which are also fainter (green), in the brightness range where \Gaia is best calibrated.}
\end{figure}

\begin{figure}[t] 
\begin{center}
\includegraphics[width=0.6\textwidth]{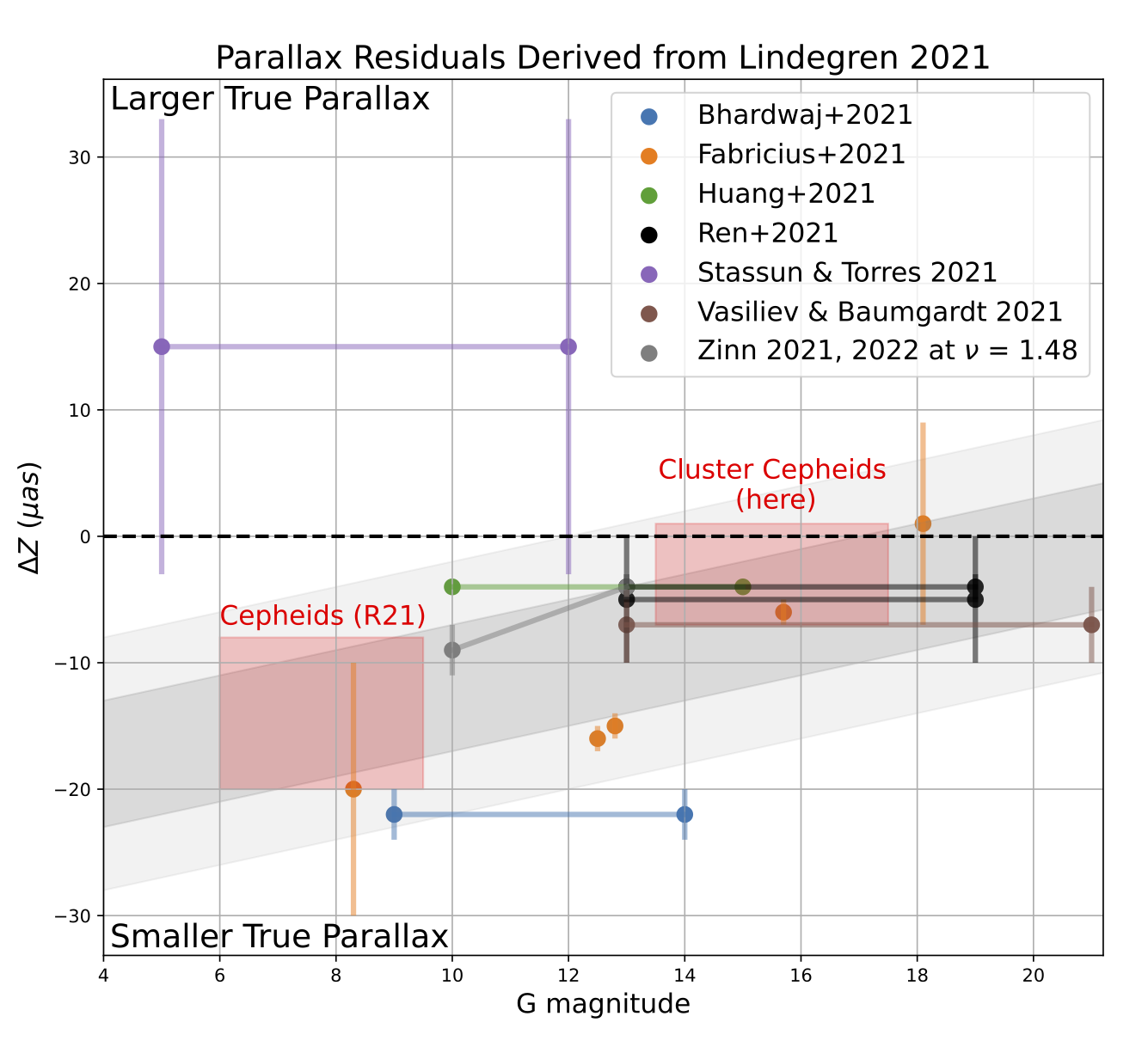}
\end{center}
\caption{\label{fg:offset} External calibrations of the residual \Gaia parallax offset relative to L21 measured by external sources as a function of source magnitude, collated by L21. We added to this the measurement from Cluster Cepheids and highlight both sets of Cepheid measurements in red. The gray regions at $\Delta Z=(G-22)$ $\mu$as with bands of $\pm$ 5 and 10 $\mu$as shows a simplistic, approximate 68\% and 95\% confidence intervals representing a ``consensus'' of the external measurements tabulated in Lindegren (2021). Measures in the negative direction imply parallaxes which are overestimated by L21.  Points from Zinn (2021,2022) are at $\nu$=1.48, the mean color of the Cepheids.}.
\end{figure}

The other important advantage of Cepheid cluster parallaxes from \Gaia comes from the better initial calibration of their parallax offset.  This term was determined for EDR3 by L21 from a combination of a million QSOs ($G>13$~mag), millions of LMC stars ($G>12$~mag), and a smaller number ($\sim$ 7000) of physically bound pairs with a bright companion ($G>6$~mag) as seen in Fig.~\ref{fg:hist}. Because the offset is a function of magnitude and was calibrated from sources that match the $12 < G < 19$~mag range of cluster stars as seen in Fig.~\ref{fg:hist}, the offset calibration for these parallaxes is optimal. The offset is less well constrained at $G<9$~mag (typical of Milky Way Cepheids) due to the smaller number of calibrating sources, requiring the {\it simultaneous} constraint of the offset at this brighter range and the Cepheid luminosity. This doubles the uncertainty in the latter, as described in \citet[][hereafter R18a]{Riess:2018a} and in R21.. Fig.~\ref{fg:offset} shows constraints on the residual (from L21) parallax offset as a function of source magnitude as determined by external studies. As expected, there is a good consensus for little to no residual offset, $zp$, at G$>13$ ($|zp|<5\mu$as).

While the \Gaia EDR3 parallaxes of cluster Cepheids would support the target $\sim$1\% goal, reaching it also requires these Cepheids to be observed on the same {\it HST} photometric system as those in SN Ia hosts, as systematic disparities (i.e., between the ground and space in the NIR), empirically seen to be $0.03-0.04$~mag \citep[][hereafter R18b and R19a]{Riess:2018b, Riess:2019a}, would otherwise dominate the other error terms. While we could estimate these Cepheid magnitudes in the 3-band {\it HST} system synthetically by applying transformations using ground-based photometry, B20 showed that the statistical errors from transformations are a factor of 3 worse than was achieved with {\it HST} (0.06 mag from the ground vs 0.02 mag per Cepheid with {\it HST}, surpassing the mean of 0.05 mag distances errors of the EDR3 parallaxes) and the additional systematic uncertainty of 0.04 mag between ground and {\it HST} zeropoints would still limit the resulting calibration of $H_0$ to $> 2\%$.

Here we present observations of 17 Cluster Cepheids measured with {\it HST} using the same instrument and three filters as other Cepheids on the SH0ES team distance ladder obtained with the same spatial-scanning protocols used in R18b and R21. In \S 2 we present the measurements, in \S 3 we use these data to constrain the Cepheid \PL relation, the scale of the distance ladder and $H_0$ with discussion of the results in \S 4.

\section{Data}

\subsection{Clusters}

The Cepheid cluster sample was initially selected as the set of 13 fundamental-mode Cepheids (including a beat or double mode Cepheid V367-SCT using its the period of its fundamental pulsation) with $P\,>4$d identified by B20 to be in clusters with high confidence. These associations were determined by comparing angular positions, proper motions, parallaxes, and ages between the \citet{Ripepi:2019} Cepheid catalogue and the open cluster catalogue derived from \Gaia DR2 by \citet{Cantat:2018}. Five of these Cepheids (DL Cas, U Sgr, XZ Car, and S Nor) were previously photometrically measured with {\it HST} by R18b and R21. The other nine were targeted in a Cycle 29 {\it HST} program, GO 16676 (PI Riess) to collect this cluster sample and are presented here for the first time.

Subsequently, \citet{Breuval:2021,Zhou:2021} and \citet[][hereafter CR22]{CruzReyes:2022} identified a few additional Cepheids hosted by open clusters using \Gaia EDR3. Three of these, WX Pup, SV Vul, and X Pup were previously measured with {\it HST} by R18b and one, X Vul, was targeted for new observations in the Cycle 29 program. Thus the full sample we consider is 17 Cepheid and cluster pairs.

From the most recent studies which use \Gaia EDR3, there is strong agreement between studies on the association of a specific Cepheid and cluster for 14 of the 17 variables we consider, which we will refer to as the ``Gold'' sample. The remaining 3 objects (which we call the ``Silver'' sample)  warrant additional consideration due to their greater spatial separation from the cluster centers, which would place them outside the cores but inside the cluster coronae which extend up to 10$\times$ farther in radius \citep{Kharchenko:2005}.

These objects are:

{\it WX Pup} -- \noindent located $\sim26\arcmin$ from the center of UBC 231, though all other properties (radial separation, proper motion, age) are in concordance. 

{\it X Pup} --\noindent  CR22 identified a new open cluster association, CL X Pup, which is $\sim30\arcmin$ from this variable with otherwise consistent measures. 

{\it XZ Car} -- \noindent B20 associated this variable with NGC$\,$3496 with a large angular separation from the core; $52\arcmin$ (or 35 pc) placing it plausibly in the clusters' corona. All other measures are consistent. \footnote{CR22 associate XZ Car with lower-than-full confidence (Silver) with a previously-undiscovered cluster, CL XZ Car.  This newly identified cluster is still far from XZ Car at $20\arcmin$ (or 14 pc), though half the separation as  NGC$\,$3496.  However, this cluster is also less rich than NGC$\,$3496, likely equalizing the difference in association advantage.  Further the new cluster is more discrepant in proper motion and photometric parallax, the latter implying a $\sim$ 250 $\pm$ 70 pc radial separation.  For these reasons we retain the B20 association though we assign this association a Silver category confidence.}

\begin{deluxetable*}{llrrrrrrrr}[b]
\tabletypesize{\footnotesize}
\tablewidth{0pt}
\tablenum{1}
\tablecaption{Cluster Data for Cepheid Hosts\label{tb:clust}}
\tablehead{\colhead{Cluster} & \colhead{Cepheid} & \colhead{RA} & \colhead{Dec} & \colhead{N} & \colhead{Size} & \colhead{$\pi\,^a$} &  \multicolumn{3}{c}{$\sigma\,^b$}\\[-6pt]
 & & \multicolumn{2}{c}{[$^\circ$]} & & \multicolumn{1}{c}{[$\arcmin$]} & \multicolumn{1}{c}{[$\mu$as]} & \colhead{stat} &  \colhead{ac} & \colhead{tot}}
\startdata
\hline
Berkeley 58 & CG-CAS &  0.0579  &  60.9421  &   206  &  5.3  &  335.9  &  2.0  &  8.2  &  8.4  \\
NGC 129 & DL-CAS &  7.5629  &  60.2252  &   469  &  16.3  &  558.6  &  1.0  &  7.0  &  7.1  \\
FSR 0951 & RS-ORI &  95.5796  &  14.6391  &   240  &  13.9  &  594.0  &  1.8  &  7.0  &  7.2  \\
vdBergh 1 & CV-MON &  99.2807  &  3.0779  &    66  &  3.4  &  578.6  &  4.2  &  8.8  &  9.7  \\
CL X-PUP$^d$ & X-PUP &  113.3532 & -20.4798  &    125  &  $<5$  &  363.0  &   --  &  --  &  7.1  \\
UBC 231 & WX-PUP &  115.5583  &  -26.2943  &    35  &  7.8  &  355.9  &  3.5  &  7.5  &  8.2  \\
Ruprecht 79 & CS-VEL &  145.2566  &  -53.8477  &   267  &  4.5  &  274.1  &  1.5  &  8.4  &  8.5  \\
NGC 3496 & XZ-CAR &  164.8822  &  -60.3360  &   554  &  7.3  &  439.0  &  0.8  &  7.6  &  7.7  \\
NGC 5662 & V-CEN &  218.7455  &  -56.6592  &   259  &  21.3  &  1322.1  &  1.4  &  7.0  &  7.2  \\
Lynga 6 & TW-NOR &  241.2089  &  -51.9538  &   163  &  4.8  &  408.2  &  2.5  &  8.3  &  8.7  \\
NGC 6067 & V340-NOR &  243.2930  &  -54.2264  &  1234  &  9.3  &  495.5  &  0.8  &  7.3  &  7.3  \\
NGC 6087 & S-NOR &  244.7350  &  -57.9140  &   279  &  16.8  &  1056.8  &  1.5  &  7.0  &  7.2  \\
IC 4725 & U-SGR &  277.9490  &  -19.1125  &   516  &  23.0  &  1540.0  &  1.1  &  7.0  &  7.1  \\
NGC 6649 & V367-SCT &  278.3591  &  -10.3966  &   628  &  4.7  &  507.9  &  1.7  &  8.3  &  8.5  \\
UBC 130 & SV-VUL &  298.0507  &  27.4452  &    98  &  8.7  &  427.8  &  2.0  &  7.3  &  7.6  \\
UBC 129 & X-VUL &  298.9809  &  26.4709  &   254  &  44.8  &  885.8  &  1.2  &  7.1  &  7.2  \\
NGC 7790 & CF-CAS &  359.6092  &  61.2159  &   248  &  5.5  &  331.0  &  1.7  &  8.1  &  8.3  \\
\hline
\enddata
\tablecomments{$a$: Includes L20 parallax offset. $b$: Total uncertainty (in last column) is the quadrature sum of the statistical uncertainty and the angular covariance (ac) $c$: data from CR22}
\end{deluxetable*}

\subsection{Cluster Parallaxes}

Cluster parallaxes were deteremined from the average of stars identified as likely members following the procedure \citet{Cantat:2018}.  The inverse of the parallax uncertainties given in EDR3 were used to weight the mean parallaxes. We provide the properties of these Cepheids and their cluster hosts in Table~\ref{tb:clust}. The cluster star parallaxes include the L21 zeropoint offset calibration term as a function of their magnitude, color and ecliptic latitude. The statistical uncertainty derived from the mean of the cluster stars is exceedingly small, ranging from 1 to 4 $\mu$as. However, \Gaia parallaxes have an angular covariance which reduces the available precision for clustered sources. Knowledge of the angular covariance has been derived \citep{Lindegren:2021b,Maiz:2021} on small scales (with numerous independent patches of $<$ 15' where the angular covariance is seen to be $\sim$ $6^2$ $\mu$as$^2$) using the consistency of LMC stars and for larger separations on degree(s) scale using QSOs, where \citet{Maiz:2021} find a covariance of $\sim$ $7^2$ $\mu$as$^2$ at 3 degree separation (their equation 9). Because the clusters vary in size from $5-40\arcmin$, it is necessary to interpolate the angular covariance determinations between these scales\footnote{\citet{Maiz:2021} summed the angular covariance on small scales as measured independently by the LMC and QSOs in their equation (10), though this double-counts the small-angle covariance measured by each, i.e., $6^2$ $\mu$as$^2$ + $7^2$ $\mu$as$^2 \sim$ $9^2$ $\mu$as$^2$. Here we use each component for the scale it best measures, LMC at $<$ 15' separations and QSO's at $>$ degree separations and interpolate the results at intermediate scales resulting in a floor for the larger clusters of $\sim$ 7 $\mu$as}. The resulting mean parallaxes with uncertainties limited by the angular covariance function are given in Table~\ref{tb:clust}. The mean of these uncertainties is 8~$\mu$as per cluster with a minimum of 7~$\mu$as and a maximum of 10~$\mu$as.  As we show in the next section, a significantly larger parallax uncertainty per cluster is not supported by the low scatter of the $\PL$ relation.  

A somewhat smaller value of the large-scale angular covariance was found by \citet[][see their Figure 5]{Zinn:2020} where the value for {\it Kepler} stars drops to $\sim4^2$~$\mu$as$^2$ at $1^\circ$ separation. For the largest cluster, UBC 129, which has a core size of 45' this result instead of the quasar-based one would reduce the uncertainty from $7\mu$as to $5\mu$as. However for most clusters the difference would be $<1\mu$as and we adopt the larger uncertainties. The mean spatial separation of the clusters is very large, $>10^\circ$ so that the cluster-to-cluster parallax covariance is negligible.

\subsection{Cepheid Photometry}

The photometry of all Cepheids were measured in 3 filters, {\it F555W}, {\it F814W}, and {\it F160W} (hereafter, $V$, $I$ and $H$ for simplicity) on Wide Field Camera 3 using the same spatial scan procedures described in R18a,b and R21. The random-phase photometry, which is independent of filter zeropoints as it depends on magnitude differences, was corrected to mean phase following the same procedures given in R18a,b and R21 and is provided in Table~\ref{tb:phot}. The sources of the light curves used to measure the phase corrections are given in Tables~\ref{tb:ground} and \ref{tb:labels}.

We combine the bands into the same reddening-free Wesenheit index used for all recent SH0ES analyses (R22), \bq m^W_H=m_{F160W}-0.386(m_{F555W}-m_{F814W}). \eq
The reddening ratio of $0.386$ is derived from the \citet{Fitzpatrick:1999} reddening law with $R_V=3.3$. Following the SH0ES convention, we correct the $m^W_H$ magnitudes for the count rate non-linearity of WFC3 relative to faint extragalactic Cepheids \citep[0.0077 mag per dex,][]{Riess:2019b} by applying this term to the Cepheids in anchors so they are directly comparable to the (uncorrected by convention) extragalactic Cepheids in SN Ia hosts. The mean difference between the MW Cepheids and the sky-limit, a floor for count rate, is 6.3 dex for an addition of 0.048 $\pm 0.004$ mag which are added, by convention, only to the Wesenheit magnitudes in Table~\ref{tb:phot}. We also include individual [Fe/H] metallicity measurements as compiled by \cite{Groenewegen:2018} for use in the Cepheid \PL relation which span a range of $-0.01$ to +0.33 dex with a mean of +0.08 dex and a dispersion of 0.07 dex, to which we add 0.06 dex to convert from [Fe/H] to [O/H] \citep{Romaniello:2021}.

\startlongtable
\begin{deluxetable*}{lrrrrrrrrrrrrll}
\tabletypesize{\scriptsize}
\tablewidth{0pt}
\tablenum{2}
\tablecaption{Photometric Data for MW Cepheids\label{tb:phot}}
\tablehead{\colhead{Cepheid} &  \colhead{log P} & \colhead{$F555W$} & \colhead{$\sigma$}  & \colhead{$F814W$} & \colhead{$\sigma$}  & \colhead{$F160W^a$} & \colhead{$\sigma$} & \colhead{$m^{W,b}_H$} & \colhead{$\sigma$}  & \colhead{Met$^d$} & \colhead{$\pi_{R22}^c$} & \colhead{$\sigma$} & \colhead{Cluster} & \colhead{Sample}}
\startdata
\hline
\multicolumn{2}{l}{Cycle 29} \\
\hline
CF-CAS &  0.688  &  11.30  &  0.023  &  9.693  &  0.026  &  8.261  &  0.027  &  7.684  &  0.030   &  -0.01  &  308.4  &  4.3  & NGC 7790 & Gold \\
CG-CAS &  0.640  &  11.56  &  0.024  &  9.843  &  0.046  &  8.315  &  0.027  &  7.694  &  0.034   &  0.06  &  330.2  &  5.1  & Berkeley 58 & Gold \\
CS-VEL &  0.771  &  11.87  &  0.015  &  10.04  &  0.019  &  8.375  &  0.027  &  7.710  &  0.029   &  0.09  &  268.5  &  3.5  & Ruprecht 79 & Gold \\
CV-MON &  0.731  &  10.45  &  0.024  &  8.587  &  0.018  &  6.924  &  0.027  &  6.250  &  0.029   &  0.09  &  559.4  &  7.6  & vdBergh 1 & Gold \\
RS-ORI &  0.879  &  8.553  &  0.016  &  7.239  &  0.027  &  6.110  &  0.027  &  5.652  &  0.030   &  0.11  &  588.3  &  8.0  & FSR 0951 & Gold \\
TW-NOR &  1.033  &  11.84  &  0.028  &  9.193  &  0.020  &  6.877  &  0.027  &  5.900  &  0.030   &  0.27  &  415.2  &  5.8  & Lynga 6 & Gold \\
V-CEN &  0.740  &  7.006  &  0.012  &  5.802  &  0.012  &  4.723  &  0.027  &  4.311  &  0.028   &  0.12  &  1347.2  &  17.3  & NGC 5662 & Gold \\
V340-NOR &  1.064  &  8.533  &  0.024  &  7.137  &  0.018  &  5.852  &  0.027  &  5.363  &  0.029   &  0.07  &  507.1  &  6.9  & NGC 6067 & Gold \\
V367-SCT &  0.799  &  11.75  &  0.043  &  9.194  &  0.030  &  7.112  &  0.050  &  6.171  &  0.054   &  0.05  &  523.0  &  13.0  & NGC 6649 & Gold \\
X-VUL &  0.801  &  9.051  &  0.019  &  7.161  &  0.033  &  5.590  &  0.027  &  4.911  &  0.031   &  0.13  &  931.9  &  13.2  & UBC 129 & Gold \\
\hline
\multicolumn{2}{l}{Cycles 22 and 27} \\
\hline
DL-CAS &  0.903  &  9.106  &  0.019  &  7.569  &  0.022  &  6.238  &  0.018  &  5.693  &  0.021   &  0.05  &  556.3  &  5.4  & NGC 129 & Gold \\
U-SGR &  0.829  &  6.886  &  0.018  &  5.388  &  0.011  &  4.143  &  0.027  &  3.619  &  0.028   &  0.14  &  1618.0  &  21.0  & IC 4725 & Gold \\
XZ-CAR &  1.221  &  8.773  &  0.017  &  7.217  &  0.006  &  5.770  &  0.007  &  5.219  &  0.010   &  0.03  &  426.5  &  1.9  & NGC 3496 & Silver \\
S-NOR &  0.989  &  6.578  &  0.011  &  5.410  &  0.012  &  4.391  &  0.012  &  3.994  &  0.014   &  0.10  &  1067.2  &  6.7  & NGC 6087 & Gold \\
WX-PUP &  0.951  &  9.191  &  0.030  &  7.944  &  0.012  &  6.807  &  0.010  &  6.372  &  0.016   &  -0.01  &  378.2  &  2.8  & UBC 231 & Silver \\
SV-VUL &  1.653  &  7.267  &  0.047  &  5.648  &  0.033  &  4.214  &  0.027  &  3.643  &  0.035   &  0.11  &  457.3  &  7.4  & UBC 130 & Gold \\
X-PUP &  1.414  &  8.695  &  0.019  &  7.128  &  0.010  &  5.628  &  0.008  &  5.073  &  0.012   &  0.02  &  340.2  &  1.8  & CL XPUP & Silver \\
\hline
\enddata
\tablecomments{$a$: Does not include addition of $0.0075 \pm 0.006 $ mag/dex to correct CRNL for 5 to 6.5 dex between MW and extragalactic Cepheids.\\$b$: Includes addition of CRNL to allow direct comparison to extragalactic Cepheids in R22 which do not include any CRNL correction.\\$c$: $\pi_{phot}=10^{-0.2(\mu-10)}$ where $\mu=m^W_H-M^W_H$ and $M^W_H$ is the absolute Wesenheit magnitude determined from the Cepheid period and the distance scale from \cite{Riess:2022} where $b_W=-3.30$, $Z_W$=-0.22 mag/dex, $M^W_{H,1}=-5.90$ mag which results in $H_0$=73.04 \kmsmpc as discussed in the text. $d$: $[$Fe/H$]$ from \citet{Groenewegen:2018}.} 
\end{deluxetable*}

\begin{deluxetable}{lllllll}[h]
\setlength\tabcolsep{0.4cm}
\def\arraystretch{0.9}
\tablecaption{Ground Data Sources for Phase Determination\label{tb:ground}}
\tablenum{3}
\tablehead{
\colhead{Identifier} & \multicolumn5c{References$^a$} \\
\cline{2-6}
& \colhead{Phase determination} & \colhead{$V$} & \colhead{$I$} & \colhead{$J$} & \colhead{$H$}
}
\startdata
CF Cas & 3-10,13,15,24,25,33,34 & 3-10,13,15,24,25,33,34 & 34 & NA & NA \\
CG Cas & 4,5,9,13,15,22,33,34,36 & 4,5,9,13,15,22,33,36 & 22,36 & NA & NA \\
CS Vel & 16,25,33-35 & 25,33-35 & 34,35 & NA & NA \\
CV Mon & 1,3,7,9-11,13,16,20,21,31,34,35 & 3,7,9-11,13,20,21,31,34,35 & 11,34,35 & NA & NA \\
RS Ori & 1,3,20,21,31,34,35 & 3,20,21,31,34,35 & 34,35 & NA & NA \\
TW Nor & 1,25,26,33-35 & 25,26,33-35 & 34,35 & NA & NA \\
V Cen & 1,11,16,23,34 & 11,23,34 & 11,23,34 & NA & NA \\
V0340 Nor & 34,36 & 34,36 & 34,36 & NA & NA \\
V0367 Sct$^b$ & 3-6,8,9,11,12,15,16,27-31,33,34 & 3-6,8,9,11,12,15,27-31,33,34 & 34 & NA & NA \\
X Vul & 3,4,6,9,10,21,31,34 & 3,4,6,9,10,21,31,34 & 34 & NA & NA \\
\enddata
\tablecomments{$a$: labels are described in Table~\ref{tb:labels}. NA: no ground data available. $b$: Light curves were modeled with two periods.}
\end{deluxetable}

\begin{deluxetable}{llclllc}[h]
\setlength\tabcolsep{0.4cm}
\def\arraystretch{0.9}
\tablecaption{References for the Labels in Table~\ref{tb:ground}\label{tb:labels}}
\tablenum{4}
\tablehead{
\colhead{ID} & \colhead{Reference} & \colhead{Comments} & & {ID} & \colhead{Reference} & \colhead{Comments}
}
\startdata
1 & \citet{Pel1976} & McMaster & & 2 & \citet{Welch+1984} & McMaster\\
3 & \citet{Moffett+1984} & McMaster & & 4 & \citet{1992AandAT....2..107B} & McMaster\\
5 & \citet{1992AandAT....2..157B} & McMaster & & 6 & \citet{1992AandAT....2....1B} & McMaster\\
7 & \citet{1992AandAT....2...31B} & McMaster & & 8 & \citet{1992AandAT....2...43B} & McMaster\\
9 & \citet{Berdnikov1992} & McMaster  & & 10 & \citet{1993AstL...19...84B} & McMaster\\
11 & \citet{Berdnikov+1995} & McMaster  & & 12 & \citet{Berdnikov+1995} & McMaster\\
13 & \citet{1995AstL...21..308B} & McMaster  & & 14 & \citet{Szabados1981} & McMaster\\
15 & \citet{1986PZ.....22..369B} & McMaster  & & 16 & \citet{Laney+1992} & McMaster\\
17 & \citet{Barnes+1997} & McMaster  & & 18 & \citet{Szabados1977} & McMaster\\
19 & \citet{Henden1980} & McMaster  & & 20 & \citet{Szabados1991} & McMaster\\
21 & \citet{Szabados1980} & McMaster  & & 22 & \citet{Henden1996} & McMaster\\
23 & \citet{Gieren1981} & McMaster  & & 24 & \citet{Berdnikov1987} & McMaster\\
25 & \citet{Harris1980} & McMaster  & & 26 & \citet{Madore1975} & McMaster\\
27 & \citet{1994AandAT....5..317B} & McMaster  & & 28 & \citet{1995IBVS.4141....1B} & McMaster\\
29 & \citet{Madore+1975} & McMaster  & & 30 & \citet{1994IBVS.3991....1B} & McMaster\\
31 & \citet{Pojmanski1997} & ASAS  & & 32 & \citet{Alfonso-Garzon+2012} & I-OMC\\
33 & \citet{2017PASP..129j4502K} & ASAS-SN  & & 34 & \citet{Berdnikov+2000} & \\
35 & \citet{2015yCat..90410027B} &  & & 36 & This work & This work\\
\enddata
\setlength\tabcolsep{6pt}
\def\arraystretch{1}
\end{deluxetable}

\ \par

\section{Constraining the PL Relation}

Briefly, to relate Cepheid luminosity to period, it is useful to define an extinction-free distance modulus from the difference in an apparent and absolute Wesenheit magnitude, $\mu_0=m_H^W-M_H^W$. This is standardized for Cepheids using the \PL relation for the $i$th Cepheid as \bq \mu_{0,i}=m_{i,H}^W-(M_{H,1}^W+b_W\, (\log\,P_{i}-1)+Z_W \ \Delta {\rm [O/H]}_{i}) \eq where $M_{H,1}^W$ is the fiducial absolute magnitude at $\log\,P\!=\!1$ ($P\!=\!10$~days) and solar-metallicity. The parameters $b_W$ and $Z_W$ (sometimes called $\gamma$ in the literature) define the empirical relation between Cepheid period, metallicity, and luminosity.
The expected parallax, $\pi_{phot,i}$ in units of mas, is given by \bq \pi_{phot,i}=10^{-0.2(\mu_{0,i}-10)} \eq

It is instructive to first examine the relation\footnote{The conversion of Gaussian parallax estimates to distance and magnitude as plotted on the $\PL$ and are functions of their inverse, may result in biased likelihoods unless the signal-to-noise of the parallaxes is sufficiently high. An approximate result with a symmetric uncertainty can be obtained with an account of the Lutz-Kelker type bias \citep{lutz73} (e.g., the volume bias, \citet{Bailer-Jones:2021}) that is still quite accurate. This is of less concern for the cluster parallaxes since their mean SNR is very high at $>$ 50. } between $\log P$ and $M_H^W=m_H^W-(5log(\pi^{-1})+10)$ before proceeding to a more rigorous determination of parameter constraints.

\begin{figure}[b] 
\begin{center}
\includegraphics[width=0.8\textwidth]{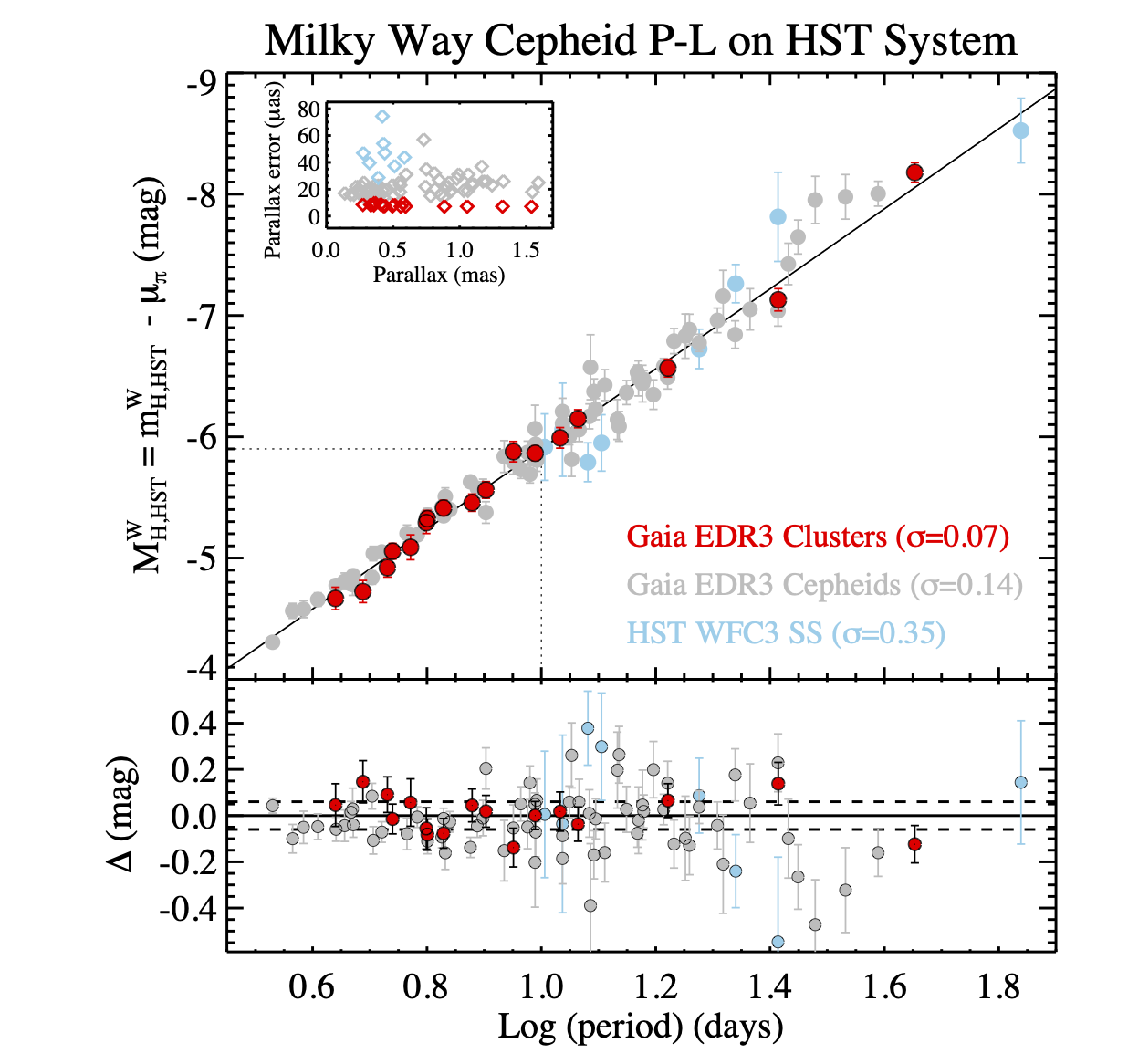}
\end{center}
\caption{\label{fg:plr} The Milky Way Cepheid period--luminosity relation in the {\it HST} NIR, reddening-free (Wesenheit) system as calibrated with three parallax samples. Parallaxes from {\it HST} spatial scanning (R18b) for 8 Cepheids are in blue and yield 3\% precision in the mean. The 66 points in gray (R21) using \Gaia EDR3 Cepheid parallaxes with simultaneous calibration of the parallax offset. The red points come from cluster Cepheids and are less dependent on parallax offset calibration as they are measured in the range where \Gaia is best calibrated. These samples differ in their parallax precision (inset) with the cluster parallaxes reaching $\sim$ 10 $\mu$as uncertainties leading to the low dispersion of $\sigma=0.07$ mag for the cluster Cepheids. The dotted line (upper) shows the reference luminosity for the distance ladder.}
\end{figure}

Fig.~\ref{fg:plr} shows the PL using this procedure; the dispersion of the Cluster Cepheids is extremely low: 0.074 mag for the Gold+Silver sample and 0.064 mag for the Gold sample. The PL parameters, $b_W$ and $Z_W$, are taken from R22. A fair share of the dispersion comes from observational errors; a mean of 0.027 mag for the Cepheid Wesenheit magnitudes and a mean of 0.044 mag from the cluster parallax uncertainties, for a quadrature sum of 0.051 mag. The remaining noise term is the intrinsic width of the NIR Wesenheit instability strip which can be estimated empirically from the excess dispersion. R19 found this to be $\sim$ 0.06 mag for 70 LMC Cepheids. The MW data support yield the same intrinsic scatter for the full sample and 0.045 mag for the Gold sample.  These low values of the intrinsic scatter which match that of the LMC (where there is negligible variations in individual Cepheid distances) provide little room to accommodate the presence of additional uncertainty in the EDR3 cluster parallaxes in Table 1.  This measurement thus provides a strong limit on unrecognized Gaia systematic errors.

We now proceed to measure constraints on parametric terms of the \PL relation following the same formalism used by the SH0ES team in R18a,b and R21, fitting the data in parallax space with the provided Gaussian parallax errors. 

We first optimize the value of: 
\bq \chi^2=\sum {{( \pi_{{\rm EDR3},i} - \pi_{{\rm phot},i} + {\it zp})^2}{\sigma_i^{-2}}}, \eq 
 where {\it zp} is a residual parallax offset {\it after} application of the L21-derived parallax offset and $\pi_{phot,i}$ depends on the Cepheid \PL parameters in equations 2 and 3, $Z_W$,$b_W$, and $M_{H,1}^W$. The individual $\sigma_i$ are derived by adding in quadrature the photometric parallax uncertainty, the intrinsic width of the NIR Wesenheit $P-L$ and the parallax uncertainties in Table 1.  For fits which can accommodate inaccuracies in the L21 zeropoint offset calibration we assign a value for $\sigma_{zp}$ for L21 as given below.
 
 In Table~\ref{tb:fits} we give the results of these fits. R21 assigned a fairly large uncertainty prior of $\sigma_{zp}=10 \mu$as due to the modest calibration sample available to L21 at this high brightness range ($6<G<10$). In constrast (see Fig.~\ref{fg:hist}), the magnitudes of the stars used to measure the cluster parallaxes well match the bulk of those used by L21 to calibrate the offset. The mean $G$ mag is 15.5 with a dispersion of 1.9 mag. So for these fits we assign a smaller, {\it a priori} uncertainty of $\sigma_{zp}=5 \mu$as (L21 states the offset uncertainty to be ``a few'' $\mu$as in the magnitude and color range where it is well calibrated so our choice of $\sigma_{zp}$ is relatively cautious). This would correspond to a 0.016 mag systematic uncertainty for the cluster sample.
 
\begin{deluxetable}{lrlcllllr}[h]
\tablewidth{0pc}
\tablecaption{Best Fits to \Gaia EDR3 Cluster Cepheids (CC)\label{tb:fits}}
\tablenum{5}
\tabletypesize{\small}
\tablewidth{0pc}
\tablehead{\colhead{Fit$^a$} & \colhead{N} & \colhead{$M_{H,1}^W$} & \colhead{{\it zp}} & \colhead{$b_W$} & \colhead{$Z_W$} & \colhead{$\sigma_{IS}$} & \colhead {H$_0$ CC only} & \colhead {H$_0$ all anchors} \\[-3pt]
\colhead{} & & \colhead {[mag]} & \colhead {[$\mu$as]} & \multicolumn{2}{c}{[mag/dex]} & \colhead{[mag]} & \multicolumn{2}{c}{[\kmsmpc]}}
\startdata
\hline
{\bf 2, G+S} & 17 & $-5.902 \pm 0.026$ & $-3\pm4$ & $-3.299 \pm 0.015 ^b$ & $-0.217 \pm 0.046 ^b$ & 0.060 & $72.9 \pm 1.3$ & {$\bf 73.04 \pm 0.99$} \\
{\bf 2, G} & 14 & $-5.907 \pm 0.024$ & $-4\pm 4$ & $-3.299 \pm 0.015 ^b$ & $-0.217 \pm 0.046 ^b$ & 0.047 & $72.7 \pm 1.3$ & {\bf $72.98 \pm 0.99$} \\
\hline
2, G+S$^c$ & 17 & $-5.893 \pm 0.018$ & 0$^d$ & $-3.36\ \ \pm 0.07$ & $-0.217 \pm 0.046 ^b$ & 0.060 & \multicolumn{1}{c}{---} & \multicolumn{1}{c}{---} \\
2, G$^c$ & 14 & $-5.907 \pm 0.018$ & 0$^d$ & $-3.44\ \ \pm 0.08$ & $-0.217 \pm 0.046 ^b$ & 0.047 & \multicolumn{1}{c}{---} & \multicolumn{1}{c}{---} \\
\hline
1, G+S & 17 & $-5.890 \pm 0.018$& 0$^d$ & $-3.299 \pm 0.015 ^b$ & $-0.217 \pm 0.046 ^b$ & 0.060 & $73.3 \pm 1.1$ & $73.16 \pm 0.97$ \\
1, G & 14 & $-5.892 \pm 0.017$ & 0$^d$ & $-3.299 \pm 0.015 ^b$ & $-0.217 \pm 0.045 ^b$ & 0.047 & $73.2 \pm 1.1$ & $73.14 \pm 0.97$ \\
\hline
\hline
\enddata
\tablecomments{{\it a}: Fit solutions are labeled by number of parameters (1 or 2) and samples (Gold, Silver); best solutions indicated with bold font. {\it b}: Fixed to R19 values. {\it c}: Cepheid luminosity not determined with same \PL parameter $b_W$ R22, so not directly applicable to determine H$_0$. {\it d}: Assuming no residual parallax offset in \Gaia EDR3.}
\end{deluxetable}

 The Cepheid cluster sample offers it's best constraints on the \PL intercept, $M_{H,1}^W$. We therefore consider the best optimizations to be 2-parameter fits for $zp$ and $M_{H,1}^W$ using constraints on the other \PL parameters ($b_W$ and $Z_W$) obtained from the external Cepheid sample (used in R22) which well constrains these parameters as given in Table~\ref{tb:fits}. 
 
 Our best fit to the full sample yields $M_{H,1}^W=-5.902 \pm 0.026$ mag and $zp=-3 \pm 4$ $\mu$as and $M_{H,1}^W=-5.907 \pm 0.024$ mag and $zp=-4 \pm 4$ $\mu$as for the Gold sample. The value of $M_{H,1}^W$ is very close to the value derived from MW Cepheid parallaxes in R21, $-5.903 \pm 0.024$ mag, and to the mean value from 3 geometric anchors (LMC DEBs, NGC 4258 masers and MW Cepheid parallaxes) of $-5.894 \pm 0.017$ mag in R22. The residual $zp$ from L21 is small or undetected, consistent with the expectation it should be small in this magnitude range where \Gaia is best calibrated. The listed uncertainty in $b_W$ and $Z_W$ is included but contributes less than 10\% of the total variance of $M_{H,1}^W$. Using this constraint as {\it the sole anchor} of the SH0ES distance ladder (R22) gives $H_0=72.9 \pm 1.3$ \kms and $H_0=73.3 \pm 1.1$ \kms with and without determination of the offset term, respectively for either sample including systematic uncertainties in the SH0ES distance ladder (see Table 5).  The Gold sample yields $H_0=72.7 \pm 1.3$ and $H_0=73.2 \pm 1.1$ with and without determination of the offset term.  
 
 Adding this constraint to the latest SH0ES measurement (R22) reduces the uncertainty in $H_0$ by $\sim$ 5\% to $H_0=73.034 \pm 0.99$ \kms with from the full sample and $H_0=72.98 \pm 0.99$ \kms for the Gold sample; simply summarized as $H_0=73.01 \pm 0.99$ \kms.  A tighter constraint on $M_{H,1}^W$ is found by fixing the parallax offset to the L21 value, i.e., $zp=0$ (and $\sigma_{zp}=0\mu$as), yielding $-5.890 \pm 0.018$ mag for the full sample and $-5.892 \pm 0.017$ for the Gold sample. This reduces the uncertainty in the luminosity by 25\% and improves the precision of $H_0$ by 7\%. The improved calibration of Cepheids in the R22 distance ladder results in a $\sim$ 5.3 $\sigma$ Hubble Tension with Planck+$\Lambda$CDM.
 
 We also fit a different 2-parameter model by fixing $zp$ to L21 and varying the slope, $b_w$, and the luminosity. The slopes are in fair agreement with the R22 mean ($-3.30 \pm 0.02$) with $-3.36 \pm 0.07$ for the full sample and $-3.44 \pm 0.08$ for the Gold sample. The values of $M_{H,1}^W$ are little changed though not directly useful for the determination of $H_0$ since they pertain to a different slope than that used in R22.

If we double the prior uncertainty in $\sigma_{zp}$ to 10 $\mu$as we find $-5.923 \pm 0.034$ mag and $H_0=72.90 \pm 1.01$ \kms and $zp=-8 \pm 7 \mu$as (if we discard any constraint on the zeropoint from L21 the sample measures the offset to $zp=$-9 $\mu$as but without detecting a difference in $zp$ from L21). If we increase the cluster parallax errors by 30\% we find $-5.905 \pm 0.028$ mag and $H_0=73.00 \pm 0.99$ \kms. However if these larger parallax errors are compensated by a smaller empirical intrinsic scatter (0.045 mag for the full sample to get $\chi^2_{dof}=1$) the result is $-5.905 \pm 0.023$ mag and $H_0=72.98 \pm 0.98$ \kms. 

If we directly compare the 17 Cepheid and cluster parallaxes we find a mean difference of 8 $\pm$ 5 $\mu$as consistent with their independently measured, optimal zeropoint difference of 11 $\mu$as with the sense that the (uncorrected) Cepheid parallaxes are larger (shorter distances) than their clusters which the analyses attribute to a significant residual offset applying only to the bright MW Cepheids in less-well calibrated range for \Gaia.

\begin{figure}[t] 
\begin{center}
\includegraphics[width=0.9\textwidth]{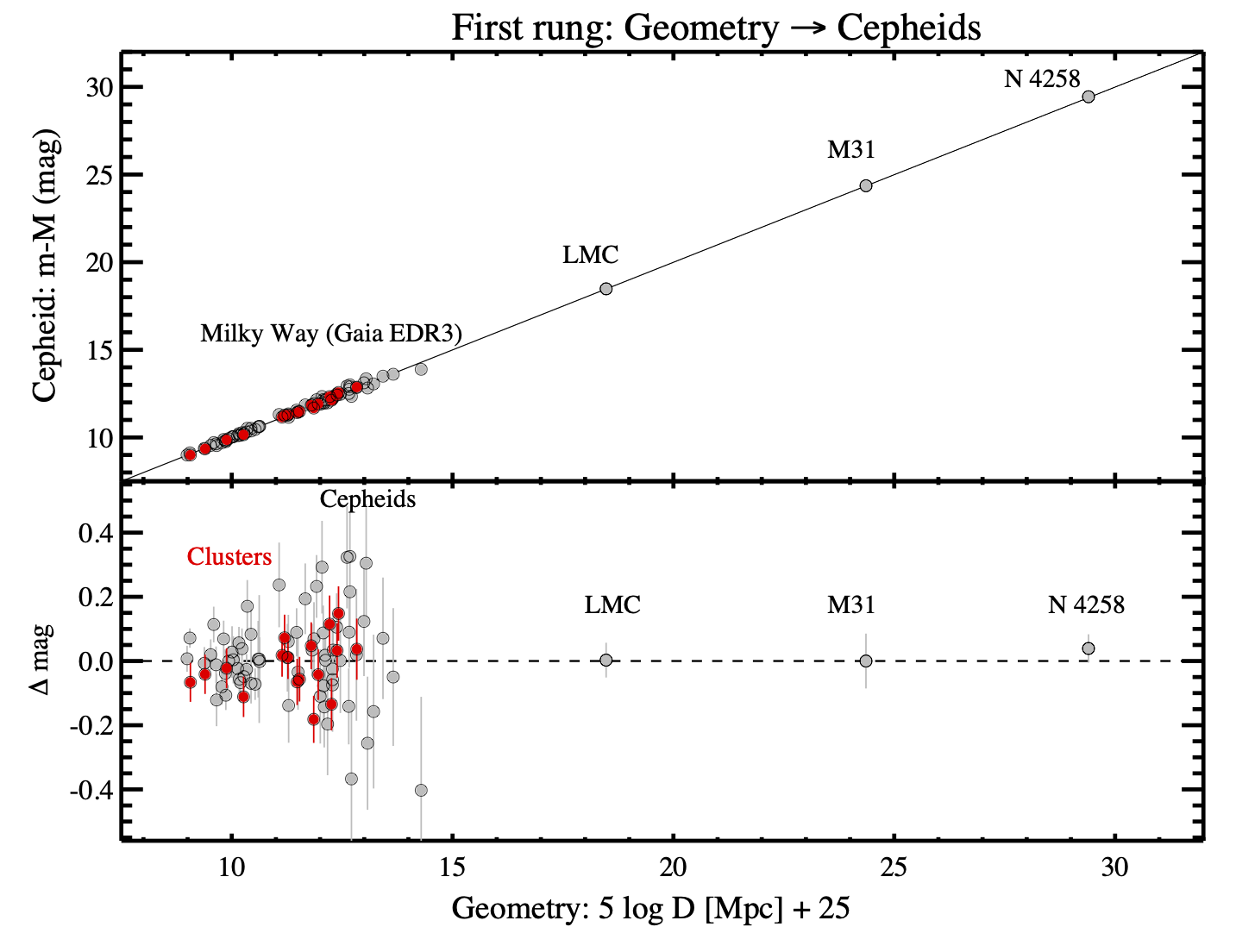}
\end{center}
\caption{\label{fg:firstrung} The first rung of the SH0ES distance ladder, geometry calibrating Cepheids.  Points in gray are as presented in R22 and points in red are the cluster Cepheids presented here.
}
\end{figure}

In summary we find a much smaller sample of Cluster Cepheids observed with {\it HST} yields a similar constraint (value and uncertainty) as the full MW {\it HST} Cepheid sample. Most importantly, this constraint is measured with a different and smaller contribution to the systematic uncertainty from the \Gaia parallax offset. In this way the two samples are complementary, demonstrating consistency in the Cepheid calibration in the presence of a significant (MW Cepheids) or negligible (Cluster Cepheids) nuisance term. 
 
\clearpage
\section{Discussion}

 The MW Cepheid field and cluster samples together provide an important check on the Cepheid calibration determined from \Gaia parallaxes in light of uncertainties related to the parallax offset term. They pass this test at the present precision available increasing our confidence in the distance scale calibrated with Cepheid variables and further reduce options for evading the present ``Hubble Tension'' via unrecognized, compounded systematic errors. Fig.~\ref{fg:firstrung} shows the status of the first rung of the SH0ES Cepheid distance ladder measured on the {\it HST} photometric system comparing geometric distances to Cepheids across 20 magnitudes, a range that exceeds the measured range of SN Ia.
With future data releases, DR4 and DR5, it is reasonable to expect further improvements to the \Gaia parallaxes that will support a $<$1\% determination of $H_0$ via the Cepheid-SN Ia distance ladder. 

\Gaia EDR3 parallaxes used to directly calibrate the luminosity of another standard candle, the Tip of the Red Giant Branch (TRGB), also point in the direction of growing ``Hubble Tension''. The parallaxes of stars in the globular cluster Omega Centauri yield $\mu_0=13.57-13.60$ mag \citep{Soltis:2021,Maiz:2021,Vasiliev:2021}, which together with the extinction-corrected tip of $m_I=9.63$ \citep{Bellazzini:2001, Bono:2008} supports the fainter end of the TRGB luminosity range ($M_I \geq -4.0$) as do \Gaia EDR3 parallaxes of field Red Giants \citep{Li:2022}. While the limited precision of these calibrations still make them consistent with non-Gaia based results (e.g., \citet{Freedman:2021} at $-4.05$ and \citet{Anand:2021} at $M_I=-4.00$), they offer an important indicator as subsequent releases from $\Gaia$ offer the only route to beat the 1.5\% precision of the best alternative, the masers in NGC 4258.  As relative distance measurements from primary and secondary distance indicators continue to improve, the ultimate test of the present Hubble Tension and $\Lambda$CDM should come from the consistency or lack there of between (final) parallaxes from \Gaia and the angular size of the acoustic scale measured from the Cosmic Microwave Background from {\it Planck}, two fundamental scale measurements of a Cosmos for which there can be only one. 

RIA \& MC acknowledge support from the European Research Council (ERC) under the European Union's Horizon 2020 research and innovation programme (Grant Agreement No. 947660). RIA further acknowledges support through a Swiss National Science Foundation Eccellenza Professorial Fellowship (award PCEFP2\_194638).  We thank Siyang Li for help with the production of Figure 2.

\bibliographystyle{apj} %
\bibliography{bibdesk}
\clearpage

\end{document}